\documentclass[submission]{SciPost}
\pdfoutput=1
\usepackage{neel}
\usepackage{cleveref}
\usepackage{parskip}
\newcommand{\ttb}{\mathrm{T}\bar{\mathrm{T}}}
\begin{document}
	\begin{samepage}
		\begin{flushleft}\huge{Exploring T-Duality for Self-Dual Fields}\end{flushleft}
		\vspace{20pt}
		{\color{myviolet}\hrule height 1mm}
		
		\vspace*{10pt}
		\begin{flushleft}
		\large 	\textbf{Subhroneel Chakrabarti}$\,{}^a$ and \textbf{Madhusudhan Raman}$\,{}^b$
		\end{flushleft}

		\begin{flushleft}
			\emph{\large ${}^a$ FZU - Institute of Physics of the Czech Academy of Sciences \& CEICO\\ Na Slovance 2, Prague 182 21, Czech Republic.}
			\\ \vspace{1mm}
            \emph{\large $\,{}^b$ Department of Physics and Astrophysics\\
  University of Delhi, Delhi 110 007, India}
            \\ \vspace{1mm}
            \href{mailto:subhroneelc@gmail.com}{subhroneelc@fzu.cz, } \href{mailto:mraman@physics.du.ac.in}{mraman@physics.du.ac.in}\\
		\end{flushleft}
		
		\begin{flushright}
			\emph{Date: \today}
		\end{flushright}

		\section*{Abstract}
		{ 
			We study avatars of T-duality within Sen’s formalism for self-dual field strengths in various dimensions. This formalism is shown to naturally accommodate the T-duality relation between Type IIA/IIB theories when compactified on a circle \emph{without} the need for imposing the self-duality constraint by hand, as is usually done. We also continue our study of this formalism on two-dimensional target spacetimes and initiate its study as a worldsheet theory. In particular, we show that Sen's action provides a natural worldsheet-based understanding of twisted and asymmetrically twisted strings. Finally, we show that the $\ttb$-deformed theory of left- and right-chiral bosons described in Sen's formalism possesses a scaling limit that is related via field-theoretic T-duality to a recently studied integrable deformation of quantum mechanics. 
		}
		
	\end{samepage}
	
	\newpage
	\vspace{10pt}
	\noindent\rule{\textwidth}{1pt}
	\pagecolor{white}
	\tableofcontents\thispagestyle{fancy}
	\noindent\rule{\textwidth}{1pt}
	\vspace{10pt}

\section{Introduction}

In Lorentzian $(4k+2)$-dimensional spacetimes, real $(2k+1)$-form field strengths can satisfy self-duality conditions.\footnote{In this paper, statements made about `chiral' or `self-dual' fields will in general be applicable to both self-dual and anti-self-dual field strengths, unless explicitly mentioned otherwise.} Such chiral fields arise naturally in string theory and, even as quantum field theories, they have been the object of sustained attention for decades now. 

It has long been believed that these theories are not amenable to a manifestly Lorentz invariant Lagrangian description. Over the years there have been many attempts to remedy this state of affairs, each with its own merits and limitations. In this paper, we focus on a formulation of chiral field strengths due to Sen \citep{Sen:2015nph,Sen:2019qit} that was in turn inspired by the spacetime action of closed superstrings in string field theory \citep{Sen:2015uaa}. Sen's formalism is manifestly Lorentz invariant, is straightforwardly quantised, and naturally inherits (as a virtue of the peculiarities of its construction) all of the expected features that a low-energy limit of a string theory ought to possess. It has undergone several important and nontrivial consistency checks, perhaps most notably the reproduction of the chiral partition function via the path integral formalism
\cite{Andriolo:2021gen,Lambert:2023qgs}. Our own explorations of Sen's formalism, which form the basis for the present work, include the study of $\ttb$-deformations of two-dimensional chiral--anti-chiral boson pairs \citep{Chakrabarti:2020dhv} and perturbative loop-level analyses \citep{Chakrabarti:2022lnn}. For a partial list of other application of this formalism in various contexts see \cite{Lambert:2019diy,Andriolo:2020ykk,Gustavsson:2020ugb,Vanichchapongjaroen:2020wza,Rist:2020uaa,Chakrabarti:2022jcb,Andrianopoli:2022bzr,Phonchantuek:2023iao,Hull:2023dgp}.

% Furthermore, 

Sen's formalism comes with several unusual features not often encountered by quantum field theorists. Let us highlight three in particular: 
\begin{itemize}
    \item [(i)] The formalism contains an auxiliary field with a wrong-signed kinetic term. This field, however, populates a sector of the full quantum theory that is completely decoupled from all the physical fields, even gravity.
    \item [(ii)] The formalism couples to a curved, dynamical spacetime metric in an unusual way.
    \item [(iii)] The formalism treats the self-dual field strength as a fundamental dynamical field and does not invoke any notion of a gauge potential.
\end{itemize}
Despite these unusual features, it is now well-established that the formalism can be utilized systematically for any quantum computation that one may wish to perform. 

In this paper, we study Sen's formalism and its relation to T-duality. This is a natural question to ask for a number of reasons. First, among all the duality symmetries of string theory, T-duality is arguably the most rigorously established. The duality manifests even at the perturbative level and is seen to be a symmetry of the string field theory action as well \citep{Kugo:1992md}. Given that the Type IIB string (and supergravity) contains a self-dual RR $5$-form fields strength, it is natural to ask how the Type IIA/IIB T-duality ``works'' within Sen's formalism. Second, T-duality can be understood from the worldsheet perspective as an exchange symmetry of momentum and winding modes. Attempts to make T-duality manifest have always involved a \textit{doubling} of the worldsheet degrees of freedom and chiral bosons make a natural appearance in that context \citep{Tseytlin:1990nb,Tseytlin:1990va} (also see the recent review \citep{Berman:2013eva} and references therein). It is natural to ask, then, if one can use Sen's formalism in in two dimensions as a worldsheet theory and if so, how does it capture the T-duality symmetry? 

Finally, there is a manifestation of T-duality that is much less popular, but fascinating nonetheless: field theoretic T-duality \citep{Taylor:1996ik}. This T-duality, while admittedly a prescription, is well-defined and leads to identification of T-dual pairs of quantum field theories and leads to, for example, the connection between Chern-Simons theory, BF theory, and matrix models \citep{Ishii:2007sy} or more recently the interpretation of four-dimensional holomorphic Chern-Simons theory as T-dual to ordinary three-dimensional Chern-Simons theory \citep{Yamazaki:2019prm}. We find that a novel scaling limit applied to $\ttb$-deformed theory of chiral bosons obtained in \citep{Chakrabarti:2020dhv} leads to a QFT which is T-dual to a peculiar integrable deformation of quantum mechanics first introduced in \citep{Grassi:2018bci}. Thus, all three avatars of T-duality find a natural immersion within Sen's formalism. In passing, we note that in \citep{Lambert:2023qgs}, a \emph{T-like} duality was already discovered in Sen's formalism.

This paper is organised as follows. In \Cref{sec:sugra} we first review the standard textbook argument for type IIA/IIB T-duality in the supergravity limit. We point out that within all standard textbook derivations the self-duality constraint is imposed \emph{by hand}, rendering the derivations mildly unsatisfactory. We then introduce Sen's formalism and explain how T-duality is manifested without any manual input. We then move on to \Cref{sec:worldsheet} where we show how Sen's formalism in two dimensions possesses the usual T-duality symmetry and in addition, also possess an new and unusual \emph{T-like} duality symmetry. We then argue that Sen's action should be taken seriously as a candidate for a worldsheet matter sector CFT. Happily, we show that the Sen's formalism can be used to provide a worldsheet-based understanding of a class new worldsheet theories dubbed \textit{asymmetrically twisted strings} \citep{Jusinskas:2021bdj}. Additionally, we explain how one can also derive these worldsheet theories within the Floreanini-Jackiw formalism for chiral bosons \citep{Floreanini:1987as,Sonnenschein:1988ug}. In \Cref{sec:ttbar} we then discuss the implications of a novel scaling limit in Sen's action and its effect on the $\ttb$-deformed theory of chiral bosons. We then go on to propose a field theoretic T-duality prescription and show that under this novel scaling limit, the $\ttb$-deformed action is T-dual to an integrable deformation of quantum mechanics \citep{Grassi:2018bci}. In particular, this map yields crucial physical insights and offers a fresh alternative perspective into the quantum integrability of both the deformed quantum mechanics as well as the $\ttb$-deformed two-dimensional theory. We conclude in \Cref{sec:discussion} with a discussion of some natural directions for future research that our multi-pronged analysis of T-duality within Sen's formalism has revealed.

\newpage

\section{T-Duality in Type II Supergravity} \label{sec:sugra}
The exemplar of a theoretical terrain that features T-duality and self-dual field strengths is the T-duality relation between Type IIA and IIB supergravities. This duality is manifested when two Type II supergravity theories are dimensionally reduced on a spatial circle, resulting in identical effective theories in nine dimensions, with appropriate field redefinitions. This is common knowledge and covered in most textbooks (see, for example \citep{Ortin:2015hya}). However, it is important to note that in all textbook derivations, there is a minor caveat. On the Type IIB side, straightforward dimensional reduction of the action does not by itself reproduce the desired effective action in nine dimensions. In order to obtain the expected identification, it is necessary to manually incorporate the self-duality condition. 

This is unsatisfactory in two respects. First, the self-duality constraint being introduced by hand is unsatisfactory. Second, even within the pseudoaction formalism for Type IIB supergravity, the self-duality condition should only be imposed on-shell; it being imposed on an effective action is... slick, at best. Of course, from the underlying string theory worldsheet description we know that T-duality is an exact symmetry and the final answer, as depicted in most textbooks, is meaningful. What we wish to address here is the dubious methodology followed to reproduce the correct target space action within the supergravity limit. The fact that T-duality is a symmetry of the target space action of full string theory is also known from string field theory techniques, see \citep{Kugo:1992md}. It is therefore quite appropriate that a string field theory-inspired formalism for Type IIB supergravity would lead to a direct derivation of the desired nine-dimensional effective action. 

It is worthwhile to quickly revisit the textbook derivation to provide context for what will follow. To keep the analysis simple, throughout this section we deal with the simplest Kaluza-Klein reduction where the $10$d metric has the form
\begin{equation}
g=\left(\begin{array}{cc}
\eta_9 & 0 \\
0 & r^2
\end{array}\right) \ ,
\end{equation}
where $\eta_9$ is the $9$d Minkowski metric and we take the direction $x^9$ to be compactified on a circle with period $x^9 \sim x^9 + L$. Here, the quantity $r$ is a dimensionless number and the physical size of the extra dimension is given by $rL$. We also ignore contributions of all the other fields, bosonic and fermionic. This is acceptable since dimensional reduction of those terms is straightforward, and it suffices to focus our attention on the RR 5-form flux kinetic term.

This kinetic term in the $10$d pseudoaction looks like
\begin{equation}
    S \sim \int_{\mathrm{10d}} F^{(5)} \wedge \star F^{(5)} = \int_{\mathrm{10d}} \mathrm{d} C^{(4)} \wedge \star \mathrm{d} C^{(4)} \ .
\end{equation}
On the circle direction we mode expand $C^{(4)}$ as
\begin{equation}
    C^{(4)} = \sum_{n \in \mathbf{Z}} \exp{\left[\frac{i 2 n \pi  x^9}{L}\right]} \left(c^{(4)}_n + c^{(3)}_n \wedge \mathrm{d} x^9 \right) \ ,
\end{equation}
where it is implicit that the fields on the right-hand side all depend solely on the non-compact flat directions. The mode expansion of the field strength is easily recovered from the mode expansion of the gauge potential. In the limit where the circle is shrunk to zero size, only the zero modes survive (we drop the subscript for them) and the effective kinetic term in $9$d assumes the form
\begin{equation}
    S \xrightarrow[rL \to 0]{} S_9 \sim L \int_{\mathrm{9d}} \left( f^{(5)} \wedge \star f^{(5)} + h^{(4)} \wedge \star h^{(4)} \right)
\end{equation}
where we have defined $f^{(5)} = \mathrm{d} c^{(4)}$ and $h^{(4)} = \mathrm{d} c^{(3)}$.\footnote{With a slight abuse of notation we are expressing both the $10d$ and the $9d$ exterior derivative as $\mathrm{d}$. From the context it should be clear that on forms of which dimensional manifold the exterior derivative acts on.}

On the other hand, dimensional reduction of Type IIA leads to an effective action which contain a kinetic term for a \textit{single} $4$-form flux, but no $5$-form fluxes exist. Of course, in $9$d one can Poincar\'e dualize the $5$-form to express it in terms of another $4$-form. But now we have an extra $4$-form field in our hands. At this point, one invokes the fact that originally $F^{5}$ is supposed to be self-dual, which identifies the two $4$-form fields to be the same and therefore it matches with the answer obtained from the dimensional reduction of Type IIA.

It is our contention that obtaining the $9$d effective action in this manner is not entirely correct in this specific situation, as the initial action for Type IIB supergravity in $10$d is not entirely legitimate. Sen's formalism is specifically designed to address this issue, and offers an all-encompassing action for Type IIB supergravity that eliminates the necessity of manually implementing constraints. The anticipated outcome is that by starting with the complete $10$d action, one should be able to obtain the required effective action in $9$d \emph{sans} any extra efforts. The remaining part of this section will explicitly establish that this is in fact the case.

Dimensional reduction of Sen's action on a circle has already been studied. For example, \citep{Sen:2019qit} studied it for chiral bosons in $2$d (also see \citep{Andriolo:2021gen}) and \citep{Andriolo:2020ykk} did it for $6$d (also see \citep{Phonchantuek:2023iao}). Our analysis is a rather straightforward extension of these previous analyses, and in the remainder of this section, we will follow and extend the analysis presented in \citep{Andriolo:2020ykk}.

\subsection{Type IIB Supergravity in Sen's Formalism}
The essential point of departure from the pseudoaction formalism is to replace the kinetic term for the RR $5$-form flux with the following action
\begin{equation} \label{eq:sen_action}
    S = \frac{1}{2} \int_{\mathbb{R}^{1,9}} \mathrm{d}P \wedge \star \mathrm{d} P + 2 \int_{\mathbb{R}^{1,9}} \mathrm{d}P \wedge Q + \int_{\mathbb{R}^{1,9}} Q \wedge \mathcal{M}(Q)
\end{equation}
There are three (not necessarily independent) novel features which we would like to highlight.

\begin{itemize}
    \item The $5$-form RR field strength is traded in for two new fundamental fields, a $4$-form $P$ with a wrong-signed kinetic term, and a \textit{field strength-like} $5$-form field $Q$ which is defined to be self-dual with respect to a \emph{flat} metric, i.e. $Q = \star Q$.\footnote{The Hodge star operator $\star$ will always be defined with respect to the flat metric. When the need arises to talk about Hodge duals with respect to the curved metric, we will use the symbol $\star_g$ instead.} Note that, this continues to be true even when the physical target space metric is dynamical and non-flat. Furthermore, the field $Q$ itself is a dynamical field and should no longer be thought of as an exact form determined by a $4$-form gauge potential.
    \item All the integrals are defined over the flat Minkowski spacetime, even when there is a dynamical metric. The only information of the physical metric is encoded into the linear map $M(Q)$ which we will define shortly. Consequently, neither $P$ nor $Q$ transform as differential forms in the physical manifold and have unusual transformation properties under diffeomorphisms\footnote{In \citep{Andriolo:2020ykk}, these fields were called \textit{pseudoforms}.}. Nonetheless, one can establish that the resulting action is invariant under all diffeomorphisms \cite{Sen:2015nph}, as it should be. Note that only in the special case where the spacetime is flat does the field $Q$ behave as an ordinary differential form.
    \item A Hamiltonian analysis reveals that the resulting quantum theory neatly decomposes into the direct sum of an unphysical free field with wrong-signed kinetic term, and another field which we recognize as the physical RR $5$-form flux; the self-duality with respect to the physical metric now follows by construction. Any interaction term depends solely on $Q$, and never $P$. 
\end{itemize}
These features are rather strange at first sight. Nonetheless, one can show that it leads to a meaningful quantum theory of interacting self-dual form fields. For more details we refer the reader to the original papers \citep{Sen:2015nph,Sen:2019qit} and also to the wonderful paper \citep{Andriolo:2020ykk} whose notations and conventions we adopt.

The action in \cref{eq:sen_action} can be dimensionally reduced in the usual way. There are a couple of questions that, at this stage, might confront the astute reader who is unaware of the details of Sen's formalism. First of all, the field $Q$ is a $5$-form, and its mode expansion is going to appear very different from that of $F^{(5)}$ obtained via exterior derivative acting on mode expansion of $C^{(4)}$. It may seem puzzling that a formulation where field strength-like variables are treated as primary variables (as opposed to gauge potentials) can lead to a $9$d effective action where we do have a standard formulation in terms of a $3$-form gauge potential. Secondly, the action, while invariant under diffeomorhpisms, are constructed out of fields which do not transform under them in the usual manner. On the other hand, in the effective action all the fields that appear have standard transformation properties under general coordinate transformations. We will see precisely how these tensions are resolved.

The first step to perform the dimensional reduction is to construct the linear map $M(Q)$. There are many ways of doing this \citep{Sen:2015nph,Sen:2019qit}, and here we will restrict ourselves to outlining the construction of \citep{Andriolo:2020ykk}. The linear map is best defined with the following properties, not all of which are independent:
\begin{itemize}
    \item The map $M$ is sensitive only to the physical spacetime metric. It acts linearly on a self-dual form and maps it to an anti-self-dual form. That is, given a self-dual form $Q = \star Q$, we have
    \begin{equation}
        M(Q) = - \star M(Q) \ .
    \end{equation}
    \item The map is symmetric in the following sense: for any two self-dual forms $Q_1, Q_2$, we have 
    \begin{equation}
     Q_1 \wedge M(Q_2) = Q_2 \wedge M(Q_1)  \ .
    \end{equation}
    \item The combination $Q-M(Q)$ is self-dual with respect to the physical metric:
    \begin{equation}
     Q-M(Q) = \star_g \left[ Q-M(Q)\right] \ .   
    \end{equation}
    \item In the specific case when the physical metric is the flat metric itself, $M$ identically vanishes and in this case (and \emph{only} this case) the physical self-dual field is the field $Q$ itself.
\end{itemize}

Now let $\lambda^I_{\pm}$ be a basis for (anti)-self-dual forms, with the index $I$ running from $1, \cdots, \frac{1}{2} \binom{10}{5}$. Similarly, let $\Lambda^I_{+}$ be a basis for self-dual form with respect to the physical metric. Then locally on a coordinate patch, we can express
\begin{equation}
    \Lambda^I_+ = \mathscr{A}^I_J \lambda^J_- + \mathscr{S}^I_J \lambda^J_+ \;,
\end{equation}
for two appropriate matrices $\mathscr{A}$ and $\mathscr{S}$. Given this map between the two basis, one can construct a linear map $M$ that satisfies all of the above properties as
\begin{equation}
    M = - \mathscr{S}^{-1} \mathscr{A} \;.
\end{equation}
A natural choice for $\lambda^I_\pm$ is 
\begin{equation}
    \lambda^I_\pm = \Gamma^I \wedge \mathrm{d}x^9 \pm \star \left( \Gamma^I \wedge \mathrm{d}x^9 \right)
\end{equation}
Here $\Gamma^I$ form a basis for $4$-forms in $9$d flat spacetime. For later convenience, we will rewrite the above equation as
\begin{equation}
    \lambda^I_\pm = \Gamma^I \wedge \mathrm{d}x^9 \pm \star_9  \Gamma^I
\end{equation}
where we have denoted the $9$d Hodge star with respect to the flat metric as $\star_9$.
A natural choice for the basis with respect to the physical metric is
\begin{align}
    \Lambda^I_+ &= \Gamma^I \wedge \mathrm{d}x^9 + \star_g \left(\Gamma^I \wedge \mathrm{d}x^9 \right) \nonumber \\
    &= \Gamma^I \wedge \mathrm{d}x^9 + \frac{1}{r}\star_9 \Gamma^I
\end{align}
One therefore obtains the linear map for our choice of compactification metric to be
\begin{equation}
    M^{IJ} = - \frac{r-1}{r+1} \delta^{IJ} \;.
\end{equation}

Next step is to mode expand the fields $P$ and $Q$. This is done in usual fashion in case of the field $P$ 
\begin{equation}
    P = \sum_{n \in \mathbf{Z}} \exp{\left[\frac{i 2 n \pi  x^9}{L}\right]} \left(\mathcal{P}_n +  \, p_n \wedge \mathrm{d} x^9 \right) \;.
\end{equation}
For the self-dual field $Q$, we first mode expand it treating it like an ordinary $5$-form
\begin{equation}
    Q = \sum_{n \in \mathbf{Z}} \exp{\left[\frac{i 2 n \pi  x^9}{L}\right]} \left(\mathcal{Q}_n +  \, q_n \wedge \mathrm{d} x^9 \right) \;.
\end{equation}
But now, since $Q = \star Q$, in $9d$ this implies $\mathcal{Q}_n = - \star_9 q_n $, and therefore it is enough to treat only one of them as independent. We choose to work with the $4$-form $q_n$.

Note that while the kinetic term is an unusual one and $Q$ can no longer be thought of as an exact form, we will still have only zero modes contributing in the limit of vanishing radius. This is because the dynamics of the higher modes of $P$ are frozen --- their masses become infinite in the limit of vanishing radius --- and in turn this freezes the dynamics of the higher $Q$ modes. If, however, the kinetic terms were absent and we only had an interaction term which depends only on $Q$, all modes will contribute equally even in the limit of vanishing radius. This curious observation will play a central role in \Cref{sec:ttbar}, but is not relevant in this particular section.

In terms of the zero modes of the $10$d fields (once again, we suppress the zero mode index), the effective $9$d action can now be obtained in the usual manner. \footnote{We keep the $r$ in the gravitational coupling term and at the end of the calculation take the limit $r \to 0$.} For later convenience we choose to redefine the field $p_n \to L \, p_n$ and $q_n \to \frac{L}{2} q_n$ and express the $9$d effective action as
\begin{equation}
S_9 = L \int_{\mathbb{R}^{1,8}}\Big[-  \frac{1}{2} \mathrm{d} \mathcal{P} \wedge \star_9 \mathrm{d} \mathcal{P} - \frac{1}{2 L^2} \mathrm{d} p \wedge \star_9 \mathrm{d} p 
 + \frac{2}{L} q \wedge \mathrm{d} \mathcal{P} - \frac{2}{L^2} q \wedge \star_9 \mathrm{d} p -\frac{2}{L^2} \frac{r-1}{r+1} q \wedge \star_9 q \Big] \,.    
\end{equation}
In order to make contact with the results of dimensional reduction of Type IIA theory, we need to express the above action in terms of a $4$-form field strength determined in terms of a $3$-form gauge potential. To do this, we introduce a 
  $3$-form $C$ through a total derivative term like so:
\begin{align}
S_9 = L \int_{\mathbb{R}^{1,8}}\Big[&-  \frac{1}{2} \mathrm{d} \mathcal{P} \wedge \star_9 \mathrm{d} \mathcal{P} - \frac{1}{2 L^2} \mathrm{d} p \wedge \star_9 \mathrm{d} p 
 + \frac{2}{L} q \wedge \mathrm{d} \mathcal{P} \nonumber \\
 &- \frac{2}{L^2} q \wedge \star_9 \mathrm{d} p -\frac{2}{L^2} \frac{r-1}{r+1} q \wedge \star_9 q  + \frac{1}{L} \mathrm{d}\mathcal{P} \wedge \mathrm{d} C \Big] \,.    
\end{align}
We can now integrate out $\mathrm{d} \mathcal{P}$ to obtain the equivalent action
\begin{equation}
S_9 = L^{-1} \int_{\mathbb{R}^{1,8}}\Big[- \frac{1}{2} \mathrm{d} C \wedge \star_9 \mathrm{d} C - \frac{1}{2} \mathrm{d} p \wedge \star_9 \mathrm{d} p  - 2 q \wedge \star_9 \mathrm{d} p -\frac{2}{L^2} \frac{4r}{r+1} q \wedge \star_9 q  - 2 q \wedge \star_9 \mathrm{d} C \Big] \,.    
\end{equation}
The next step is to integrate out the field $q$. It is useful to define another $3$-form field at this stage as follows
\begin{equation}
    \tilde{C} = p + \frac{1+r}{1-r}\, C \,. 
\end{equation}
After integrating out $q$, the resulting equivalent action is
\begin{equation}
    S_9 =  \frac{1}{L} \frac{1}{1-r}\int_{\mathbb{R}^{1,8}} \mathrm{d} C \wedge \star_9 \mathrm{d} C - \frac{1}{rL} \frac{1-r}{4}\int_{\mathbb{R}^{1,8}} \mathrm{d} \tilde{C} \wedge \star_9 \mathrm{d} \tilde{C} \;.
\end{equation}
This action consists of two free $4$-form fluxes expressed in the standard manner. However, if we now take the limit $r \to 0$, we see that the flux corresponding to $C$ has the wrong sign and that corresponding to $\tilde{C}$ has the correct sign, which we identify as the desired \textit{physical} flux as obtained from dimensional reduction of type IIA. 

It is interesting to note that the kinetic term for the physical flux scales as $\sim \,(rL)^{-1}$ which is unusual from type IIA perspective.\footnote{Recall, the physical size of the curled up dimension is given by $rL$.} The standard Kaluza-Klein reduction should give a kinetic term that scales as $\sim\, (rL)$. This unusual scaling is actually a desirable feature from a $6$d context of the worldvolume theory of multiple $M5$-branes \citep{Witten:2009at,Andriolo:2020ykk}. However, as already pointed out in \citep{Andriolo:2020ykk}, within free theory, this scaling is not conclusive. It remains true that dimensional reduction of Sen's action over a circle may lead to such a dependence in $6$d that continues to survive interactions. But in our case we know this scaling cannot be the correct one. In free theory, this is easily remedied by a further field redefinition $\mathscr{C} = (rL) \Tilde{C}$.  We know from the underlying worldsheet theory that $\mathscr{C}$ should indeed be the correct field. This conclusion should \emph{emerge automatically} if one keeps track of the interaction terms involving the parent field $Q$ and demand they match with the lower dimensional counterpart obtained from Type IIA. 

\begin{figure}[ht]
\centering
\includegraphics[width=0.75\textwidth]{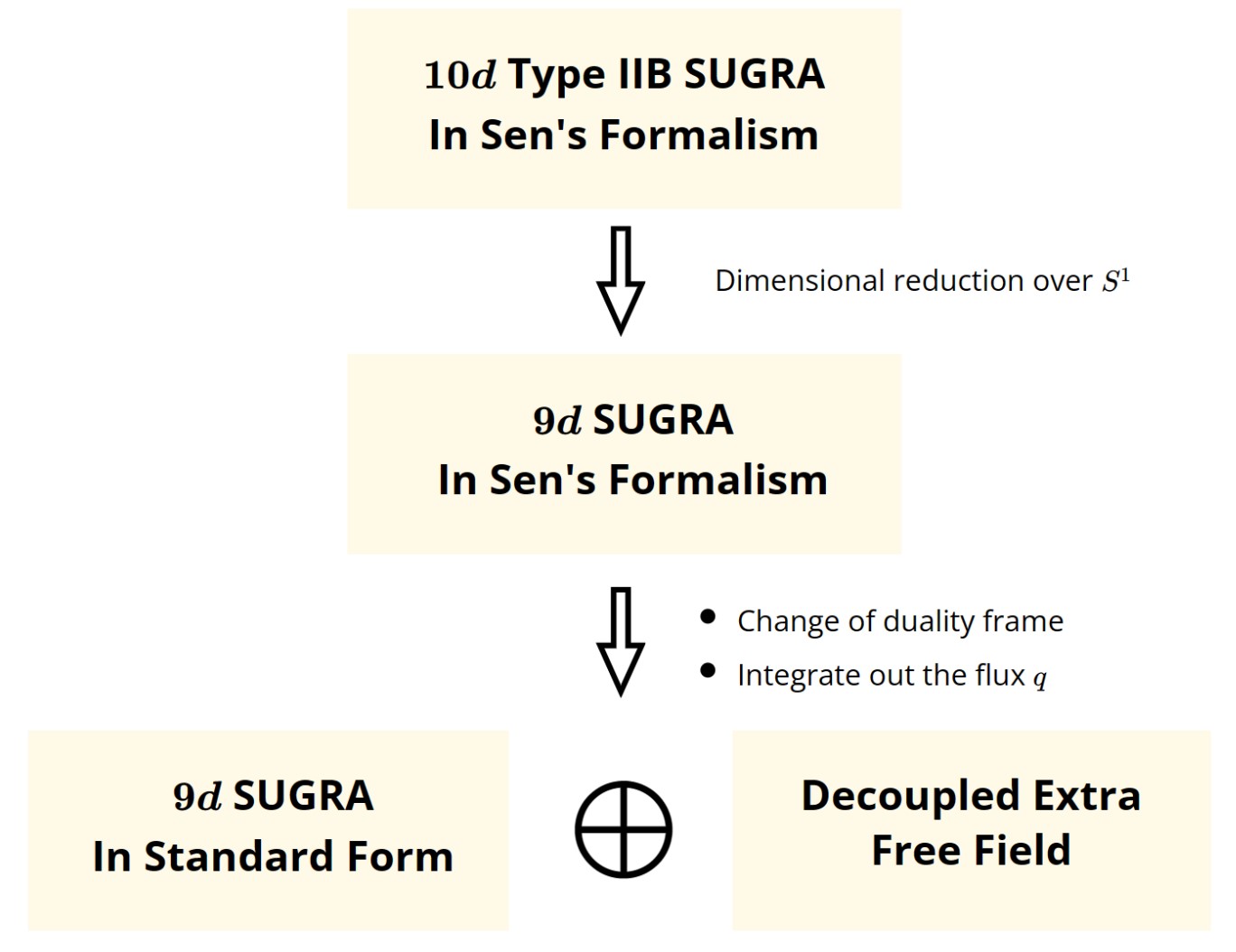}
\caption{\label{fig:typeII_T-dual}A schematic outline of how Type IIB supergravity in Sen's formalism leads to the same effective action in $9d$ as obtained from Type IIA supergravity.}
\end{figure}

While the technical steps in obtaining the effective $9$d theory from Type IIB supergravity action in Sen's formalism are rather straightforward, it reveals a couple of extremely intriguing conceptual aspects that we would like to emphasize now. First, it may seem that by using Sen's formalism, the Buscher's T-duality rules between $10$d fields are now rather convoluted (see \Cref{fig:typeII_T-dual} for a schematic summary). After all, there is no straightforward map from $Q$ to RR $4$-form flux in Type IIA at the level of the action itself. We would argue this is to be expected since such a natural map should exist only after we re-express Type IIA supergravity in Sen's formalism as well. This might seem an absurd exercise, since we have a perfectly well-defined action for Type IIA already. Recall, however, that the original motivation in Sen's formalism was Type II string field theory \citep{Sen:2015uaa}, where the unusual formalism that leads to the kinetic term for the RR $5$-form flux is also responsible for all RR sector fields in both Type IIA and IIB theories. Furthermore, as already pointed out in the original reference \citep{Sen:2015nph} (and indeed in any standard review of superstring perturbation theory, see \citep{Witten:2012bh} for example), even in the RNS formulation of the worldsheet theory, the RR sector vertex operators for \textit{all} RR-fluxes (in the canonical picture) are solely expressed in terms of field strength \textit{without} invoking any gauge potential. Furthermore, the field theoretic description of the RR fluxes that match with results from string perturbation theory must necessarily involve \textit{only} the RR field strengths. It stands to reason that a rewriting of Type IIA supergravity in Sen's formalism, while not strictly necessary, should lead to a set-up where its genesis from string theory is more manifest. Within such a reformulation, we expect that the Buscher rules would assume a very natural form. Furthermore, we will also contend that since string field theory is manifestly T-duality invariant \citep{Kugo:1992md}, a supergravity that is truly obtained as a Wilsonian effective action from string field theory action \citep{Sen:2016qap} should possess manifest T-duality \textit{and} where Sen's formalism is used to express \textit{all} of the RR fluxes. 

Second, in $10$d, the physical RR $5$-form flux is identified to be the combination $Q -M(Q)$. In the $9$d theory, in Sen-like variables, the physical field strength is indeed just $q$, which, following our definition, is simply proportional to $\mathrm{d}\Tilde{C}$. However, if one examines the composition of the gauge potential $\Tilde{C}$, one finds that it is built out of components descending from both $P$ \textit{and} $Q$ in the parent $10$d theory, namely $p$ and $C$.\footnote{Note that while $C$ was introduced to dualise the variable $\mathrm{d}\mathcal{P}$, the procedure of integrating out $\mathrm{d}\mathcal{P}$ makes $C$ cognizant of both $q$ and $\mathcal{P}$.} This is fascinating since it strongly resonates with the observation first presented in \citep{Lambert:2023qgs} that the \textit{extra} field $P$ seems to play a much more important role than just allowing us to construct a covariant kinetic term for a self-dual field strength. This also suggests that in going from the standard formulation of Type IIA theories to a Sen-like formulation, the \textit{unphysical} gauge potentials should also encode the information regarding the \textit{extra} field such that both formulations agree on the physical field strength. It will be fascinating to construct such a Sen-like action for Type IIA supergravity from the bottom up and better understand the role of the \textit{extra} field.

\section{T-Duality on the Worldsheet} \label{sec:worldsheet}

In this and the subsequent sections, we will now specialise to the
case of chiral bosons in two dimensions. Its free field dynamics in
flat spacetime is specified within Sen's formalism by the Lagrangian
\begin{equation}
  \label{eq:sen-formalism-free-action}
L _{A} = \frac{1}{2} \, \mathrm{d} \phi \wedge \star \mathrm{d} \phi - 2 A \wedge \mathrm{d} \phi \ .
\end{equation}
Here, we will use the symbol $ A $ to denote a self-dual $ 1 $-form
and $ \phi $ the auxiliary scalar field with a kinetic term that has
the wrong sign. Similarly, the dynamics of an anti-self-dual
$ 1 $-form $ B $ is governed by the Lagrangian:
\begin{equation}
L _{B}  = \frac{1}{2} \mathrm{d} \psi \wedge \star \mathrm{d} \psi + 2 B \wedge \mathrm{d} \psi \ ,
\end{equation}
where $ \psi $ is the auxiliary scalar field.

We discuss here an implementation of the standard T-duality procedure
for compact bosons adapted to the case of Sen's formalism. Usually, for a compact boson $\varphi$ with a standard kinetic term and radius $\beta$, one proceeds by treating $\mathrm{d} \varphi$ as a variable and imposes the requirement of integer winding number using a second compact boson $\mathrm{d} \tilde{\varphi}$ as a Lagrange multiplier. One then integrates out $\varphi$ and finds an action for a compact boson $\tilde{\varphi}$ with dual radius $\tilde{\beta} = \left(2\pi\beta\right)^{-1}$. Carrying out an analogue of this procedure in position space is made difficult by the non-standard kinetic term for the auxiliary scalar field $\phi$ in \cref{eq:sen-formalism-free-action}, but fortunately there is a way out of this difficulty: to work in momentum space. 

Our
strategy will be to start with the momentum space action corresponding to \cref{eq:sen-formalism-free-action} and
\begin{itemize}
\item [(i)] integrate out the auxiliary scalar $ \phi  $,
\item [(ii)] impose flux quantisation on the self-dual $ 1 $-form $ A $ via
  the Lagrange multiplier $ \tilde{B} $,
\item [(iii)] integrate out the $ 1 $-form $ A $ to get a quadratic generating
  functional for $ \tilde{B} $, and
\item [(iv)] reintroduce an auxiliary scalar $ \tilde{\psi } $.
\end{itemize}
This procedure, as we will shortly see, exchanges Sen's Lagrangian for a self-dual $ 1 $-form
$ A $ and an auxiliary scalar $ \phi $ with an \emph{anti}-self-dual
$ 1 $-form $ \tilde{B} $ and its auxiliary scalar $ \tilde{\psi }
$.

The result of (i) was written down explicitly in
\cite{Chakrabarti:2022lnn}, and we reproduce it below:
\begin{equation}
S = \frac{1}{2} \int _{} ^{} \frac{\mathrm{d} ^{2}k}{\left( 2 \pi  \right)^{2}} \left[ A _{\rho }(k) \, \mathcal{P}^{\rho \mu  }_{-} \, \Pi _{\mu \nu }(k) \, \mathcal{P}^{\nu \sigma }_{+} \, A _{\sigma }(-k) \right] \ ,
\end{equation}
where $ \mathcal{P}_{\pm } $ are projectors onto the space of
self-dual $ (+) $ and anti-self-dual $ (-) $ $ 1 $-forms
respectively
\begin{equation}
\mathcal{P}^{\mu \nu }_{\pm } = \frac{1}{2} \left( \eta ^{\mu \nu }\pm \epsilon ^{\mu \nu } \right) \ ,
\end{equation}
and the propagator for self-dual $ 1 $-forms in momentum space is
\begin{equation}
\Pi _{\mu \nu }(k) = -4 \frac{k _{\mu }k _{\nu }}{k ^{2}} \ .
\end{equation}
To make the effects of T-duality explicit, we will introduce the
dimensionless radius $ r $ of the compact chiral boson. On suppressing
indices to lighten the notation, we have
\begin{equation}
S _{A} = \frac{r ^{2}}{2} \int _{} ^{} \frac{\mathrm{d} ^{2}k}{\left( 2 \pi  \right)^{2}} \left[ A \cdot  \mathcal{P}_{-}  \Pi  \mathcal{P}_{+} \cdot A \right] \ ,
\end{equation}
Now, we wish to impose a flux quantisation condition of the form
\begin{equation}
\frac{1}{2 \pi } \oint A \in \mathbb{Z} \ ,
\end{equation}
which essentially tells us that our self-dual $ 1 $-form is
compact. To impose this condition, we introduce a second compact
anti-self-dual $ 1 $-form $ \tilde{B} $ and add to the action a term
of the form
\begin{equation}
\frac{1}{2 \pi } \int _{} ^{} A \wedge \star \tilde{B} \ .
\end{equation}
Now we do the path integral over the $ 1 $-form $ A $ which in turn
requires that we invert the kinetic term. As discussed in
\cite{Sen:2015nph}, this kinetic operator --- understood as a map from
the space of self-dual $ 1 $-forms to the space of anti-self-dual
$ 1 $-forms --- does admit a formal inverse, viz.~a map from the space
of anti-self-dual $ 1 $-forms to the space of self-dual $ 1
$-forms. This requires only that we swap the projectors flanking the
propagator. After doing the path integral, we get
\begin{equation}
  S _{\tilde{B}} = \frac{\tilde{r}^{2}}{2}  \int _{} ^{} \frac{\mathrm{d} ^{2}k}{(2 \pi )^{2}} \left[ \tilde{B} \cdot \mathcal{P}_{+}\Pi \mathcal{P}_{-} \cdot \tilde{B} \right] \ ,
\end{equation}
where we define the dual radius $ \tilde{r} $ of the compact
anti-self-dual $ 1 $-form $ \tilde{B} $ as
\begin{equation}
  \label{eq:dual-radius}
\tilde{r} = \frac{1}{2 \pi r } \ .
\end{equation}
Reintroducing the auxiliary scalar $ \tilde{\psi } $ and transforming back to position space to recover Sen's
Lagrangian for an anti-self-dual $ 1 $-form is at this point
straightforward. To summarise: the sequence of transformations we just
described sends a compact, self-dual $ 1 $-form $ A $ with radius
$ r $ to a compact, anti-self-dual $ 1 $-form $ \tilde{B} $ with dual
radius $ \tilde{r} $ given by \cref{eq:dual-radius}.

From this observation, it follows that a pair of compact self-dual
$ (A) $ and anti-self-dual $ (B) $ $ 1 $-forms with radius $ r $,
under the above sequence of transformations, is mapped into itself,
i.e.~a pair of compact anti-self-dual $ (\tilde{B}) $ and self-dual
$ (\tilde{A}) $ $ 1 $-forms with dual radius $ \tilde{r} $. This is
what worldsheet T-duality looks like in Sen's formalism. In the following sections, we will refer to the theory with both chiral and anti-chiral bosons as the $AB$ theory.

In closing this section, we mention that we have in this section focused on the effect T-duality has on the physical chiral fields. One could just as easily ``dualise'' the auxiliary scalar $\phi$ by gauging the shift symmetry $\phi\rightarrow\phi + c$, then integrating out the field $\phi$ to get an effective action for the Lagrange multiplier field coupled to the chiral field. In fact, such a T-duality exists even for just the chiral sector itself. Since the additional field is unphysical, it is \textit{a priori} unclear if this T-duality has any physical interpretation or utility. However, note that \citep{Lambert:2023qgs} it was noted that an electromagnetic duality of the extra field had physical consequences for the chiral field. So it is possible this version of T-duality, while not amenable to standard worldsheet interpretation, can have consequences in a different guise.

\subsection{Sen's Action as a Worldsheet Theory}

It is now evident that similar to the free boson action in worldsheet string theory, pairs of chiral and anti-chiral bosons in Sen's formalism display target space duality. This prompts us to investigate the possibility of understanding $AB$ theory as the matter sector of a worldsheet theory. There are a few instances in the literature where chiral bosons naturally appear within worldsheet theory. The worldsheet action of heterotic strings is the earliest and perhaps the most well-known example. Because the worldsheet theory is an on-shell formalism, the action of the chiral bosons is usually not a concern for describing heterotic strings. Nonetheless, they do become significant in some circumstances. For example, to establish the precise connection between heterotic strings and $M5$-branes wrapped on a $K3$ surface, \citep{Cherkis:1997bx} needed to incorporate chiral bosons in the action of the heterotic worldsheet by making use of the Floreanini-Jackiw formalism \citep{Floreanini:1987as,Sonnenschein:1988ug}. 

The arena of the doubled worldsheet theories is another setting where chiral bosons occur naturally (\citep{Tseytlin:1990nb,Tseytlin:1990va}, also see the recent review \citep{Berman:2013eva} and references therein). The principal goal of these models is to make T-duality manifest from a worldsheet perspective. Typically, in most constructions of doubled worldsheets, either the Floreanini-Jackiw or the Pasti-Sorokin-Tonin approach \citep{Pasti:1996vs} is adopted. Sen's formalism in both of these cases should be equally relevant. It would be an interesting exercise to test whether or if it may provide new insight into these worldsheet theories. 

More interestingly, there is a third, and a comparatively recent class of worldsheet models where chiral bosons actually play a very useful role, which to the best of our knowledge has not been pointed out so far. This is the arena of twisted and asymmetrically twisted strings \citep{Jusinskas:2021bdj}. In the remainder of this section, we quickly describe how to proceed to standard worldsheet string theory from Sen's action, and then, as a concrete example, show how to use it to give an underlying worldsheet theory explanation for asymmetrically twisted strings introduced in \citep{Jusinskas:2021bdj}.

For definiteness, we work with the $AB$ theory. The generalization to an arbitrary number of pairs of chiral and anti-chiral bosons is straightforward. For real fields, the chirality condition in $2$d can only be imposed only if the worldsheet is Lorentzian. For Euclidean theories, there is unfortunately no natural analytic continuation since the off-shell self-duality condition is not amenable to a Wick rotation. However, on-shell, the space of solutions can naturally be analytically continued, as we will soon show, and for all standard worldsheet computation this is enough. 

On-shell, a chiral boson in Sen's formalism is captured solely by the component $A_-$ which satisfies the equation of motion $\partial_+ A_- = 0$.\footnote{An appendix in \citep{Chakrabarti:2022lnn} lists the conventions we work with.} In the remainder of this section we will drop the lightcone index and only work with this non-vanishing component. We can now easily analytically continue the on-shell objects to a corresponding holomorphic Euclidean primary of scaling dimension $1$ (i.e. it transforms like a holomorphic current). Furthermore, given the $2$-point functions of these fields \citep{Chakrabarti:2022lnn}, we can also determine the OPE structure of this Euclidean field:
\begin{equation}
    A(x^-) \xrightarrow{\text{Wick}} A(z) \quad \mathrm{and} \quad
    A(z) A(w) \sim \frac{1}{(z-w)^2} \;.
\end{equation}
Furthermore, the Lorentzian stress tensor \citep{Chakrabarti:2020dhv} of the free theory now maps to a purely holomorphic stress tensor
\begin{equation}
    T(z) = \frac{1}{2} :A(z) A(z): \;.
\end{equation}
It is easy to verify that the stress tensor satisfies the expected OPE of a CFT with a holomorphic central charge $c =1$. An exactly analogous story holds true for the anti-chiral field $B_+$, which now leads to an antiholomorphic field $B(\Bar{z})$ and the resulting theory defines an antiholomorphic CFT with central charge $\bar{c}=1$ as well. The pair therefore has exactly the same behavior as the free boson CFT. The only ingredient missing from the repertoire of standard worldsheet CFT is a candidate for the vertex operator.

To do this, let us write the Laurent series expansion for the two fields as follows.
\begin{equation}
    A(z) = \sum_{n \in \mathbb{Z}} \frac{A_n}{z^{\,n+1}} \quad, \quad B(\bar{z}) = \sum_{n \in \mathbb{Z}} \frac{B_n}{\bar{z}^{\,n+1}}
\end{equation}
We first define the function $\Tilde{X}(z,\Bar{z})$ formally through the following Laurent expansion  
\begin{equation} \label{eq:X-mode_1}
    \Tilde{X}(z, \bar{z})= - i\left(A_0 \ln z + B_0 \ln \bar{z}\right)+i \sum_{n \neq 0} \frac{1}{n}\left(A_n z^{-n} + B_n \bar{z}^{-n}\right)\;.
\end{equation}
 The function $\Tilde{X}$ is not quite enough to be interpreted as a target space co-ordinate. We would ideally want to identify $A \sim \partial X$ and $B \sim \Bar{\partial}X$ at this point for some worldsheet scalar $X$. But two interrelated ingredients are missing. Essentially, dynamics of a string is perfectly captured by free bosons, and that differs from a pair of chiral bosons by a very simple, but crucial ingredient - the presence of the zero mode. Our $AB$ system is blissfully unaware of any zero mode. In fact, we can realize that the \cref{eq:X-mode_1} is essentially saying

 \begin{equation}
     i \Tilde{X}(z, \bar{z}) = \int \mathrm{d}z \, A(z) + \int \mathrm{d} \Bar{z} \, B(\Bar{z}) \;.
 \end{equation}

 We can think of the zero mode entering via the integration constant. There are two unrelated integrals, and therefore \textit{a priori} it can introduce two different integration constants. However, what matters in the end is the sum which we identify as the zero mode of the target space co-ordinate function, viz. $x_0$. Writing this \textit{integration constant} explicitly, we define the co-ordinate function as

 \begin{equation}
    X(z, \bar{z}) = x_0 + \Tilde{X}(z, \bar{z}) \;.
 \end{equation}
 
 Once we have the zero mode, the conjugate momentum to $x_0$ should be now given by the average of $A_0$ and $B_0$. Furthermore, $A_0$ and $B_0$ now will inherit some properties depending on the conditions imposed on the target space co-ordinate $X$ under rotation of the worldsheet co-ordinate $z \to e^{2 i \pi}z$. Basically, for a non-compact direction of target space we must have $A_0 = B_0$ so that
 \begin{equation}
     X(e^{2 i \pi}z, e^{-2 i \pi}\bar{z}) = X(z,\Bar{z})
 \end{equation}
 and whereas for compact directions 
 \begin{align}
     X(e^{2 i \pi}z, e^{-2 i \pi}\bar{z}) &= X(z,\Bar{z}) + 2 \pi r n \;, \quad n \in \mathbb{Z}\nonumber \\
     \implies A_0 - B_0 &= r n \;.
 \end{align}
 These are the same relations one would have in ordinary free boson CFT as explained in various textbooks (see, for example \citep{Blumenhagen:2009zz}). Let us focus on the non-compact case, and denote the conjugate momentum as $\pi_0 = A_0 = B_0$. We then have the target space coordinate function along a non-compact direction defined as
 \begin{align}
     X(z, \bar{z}) &= x_0 + \Tilde{X}(z, \bar{z}) \Big\vert_{A_0 = B_0 = \pi_0} \nonumber \\
     X(z, \bar{z}) &= x_0 - i\left(\pi_0 \ln z + \pi_0 \ln \bar{z}\right)+i \sum_{n \neq 0} \frac{1}{n}\left(A_n z^{-n} + B_n \bar{z}^{-n}\right) \;.
 \end{align}
 
 One can then check that the vertex operator $:\exp{\left(i k \cdot X\right)}:$ is a conformal primary, with respect to the total stress tensor, with both holomorphic and antiholomorphic weight being equal to $\frac{k^2}{2}$. Finally, since the original theory is cognizant only of the fields $A$ and $B$, which we can now express as $i \partial X$ and $i \bar{\partial}X$ respectively, it stands to reason that the variable $X$ will enjoy a shift symmetry, i.e. $X \sim X + c$. This is enough to recover the spacetime momentum conservation in evaluating correlation function of the vertex operators. At this point the rest of the worldsheet CFT techniques can follow as usual. We supplement the matter CFT with a $(b,c)$-ghost system, which then leads to the familiar BRST current helping us in determining the physical states as its cohomology.

So far, there is no distinct advantage in choosing to work with the $AB$ system instead of a free boson. The advantage becomes clear once we realize that since the holomorphic and antiholomorphic parts have different origins, we can treat them differently right at the outset, leading to a new class of worldsheet theories. For example, let us keep the field $A$ untouched, but in the original Lorentzian theory re-scale the field in the anti-chiral sector as follows-
\begin{equation} \label{eq:scaling}
    B \to \hat{B} = \frac{1}{\sqrt{n}} B \;.
\end{equation}
The effect of this rescaling on the antiholomorphic sector is easily seen to be 
\begin{align}
    \hat{B}(\bar{z})\hat{B}(\bar{w}) &\sim \frac{1}{n(\Bar{z}-\Bar{w})^2} \\
    \hat{\Bar{T}}(\Bar{z}) &= \frac{n}{2} :\hat{B}(\Bar{z})\hat{B}(\Bar{z}): \;.
\end{align}
These modifications, with the restriction $n \in \mathbb{Z}$, of the antiholomorphic sector with an unchanged holomorphic sector have appeared before in the literature \citep{Jusinskas:2021bdj} and were dubbed \textit{asymmetrically twisted strings} (ATS). In particular, the case $n = -1$ had appeared even earlier \citep{Hohm:2013jaa}, later given the name \textit{twisted strings} and whose tensionless limit was shown to be the ambitwistor strings \citep{Siegel:2015axg}. In particular, the ATS has in its spectrum all the massless states of a closed string and the $n$-th massive states of the open string. While these theories are not expected to lead to well-defined target space quantum theories, they are expected to capture the interaction of higher spin massive multiplets with itself as well as with the massless states. As explained in the original paper \citep{Jusinskas:2021bdj}, there are many aspects of traditional string theory where ATS could potentially provide insight that is currently inaccessible by other means.

At this junction, one can argue that there is nothing special about the use of Sen's formalism to provide an underlying worldsheet theory for ATS. Indeed, what is special is the use of a pair of chiral bosons, instead of a single free boson. More concretely, in the Sugawara form of the Floreanini-Jackiw formalism \citep{Sonnenschein:1988ug}, the Sugawara current $\bar{j}(\bar{z})$ play the same role as the field $B$ and the expression of the two stress tensors are identical with the map $\Bar{j} \to B$. So, one can also obtain the ATS OPE starting with the worldsheet theory described in the Floreanini-Jackiw formalism by simply scaling $\Bar{j} \to \frac{1}{\sqrt{n}}\Bar{j}$. To the best of our knowledge, the fact that chiral bosons can be used as a worldsheet theory underlying the OPE structure of twisted and asymmetrically twisted strings has not been pointed out earlier.

Additionally, all of this construction has natural synergy with supersymmetry, both on the worldsheet and in spacetime. One can just as well adopt the $AB$-system to describe the bosonic sector of the worldsheet of a superstring. 

Note that if one performs a worldsheet T-duality for ATS described in AB theory, one ends up with a new ATS whose holomorphic sector is now twisted, but antiholomorphic sector is unchanged. It may seem naively that this leads to a different theory, and at the level of the worldsheet, it indeed does. However, in a worldsheet theory, the physical states of interest are the spacetime fields which are captured by the BRST invariant vertex operators. Following the construction of the ATS in the original paper \citep{Jusinskas:2021bdj}, it is easy to see that the target space physics of ATS is insensitive to which of the two sectors is twisted, as long as there is a relative twist. This suggests that the spacetime physics described by ATS continues to be invariant under T-duality. The massless states of the ATS are the same as that of the closed string, and therefore they are expected to exhibit some natural T-duality symmetries. What is interesting is that the spacetime effective action of ATS, which has massive states and contain an infinite order derivative terms, continues to be T-duality invariant as well (under suitable Buscher-like rules). Our analysis therefore suggests critical strings are not the only kind of strings that lead to a T-duality invariant non-local Lagrangian. This can have interesting consequences for investigations that seek to constrain stringy $\alpha'$ corrections to supergravity by the means of T-duality alone.

So far we have been working on a Lorentzian target spacetime. A natural next step is figuring out how to incorporate a curved target space metric as well as turning on other target space background fields within Sen's formalism. We believe that the analysis of this section provides compelling evidence to take the $AB$-system as a candidate matter sector for a worldsheet action and leave the work of extending it to non-trivial target spacetime to the near future.

\section{T-Duality in Field Theory} \label{sec:ttbar}

The chiral boson in \cref{eq:sen-formalism-free-action} couples to a
gravitational background in a non-standard way and is therefore a
quantum field theory with a novel articulation of diffeomorphism
invariance. (See \cite{Hull:2023dgp} for a recent discussion of these
issues.) It is therefore interesting to ask how such theories
respond to stress-tensor perturbations, in particular, the eminently
tractable $\ttb$ deformations \cite{Jiang:2019epa}, which may be
interpreted as coupling the two-dimensional theory to a random
geometry \cite{Cardy:2018sdv}. Briefly, the $\ttb$ deformation (at the classical level) is a flow on the space of two-dimensional quantum field theories that is at each step triggered by the composite operator $\mathrm{det} T$:
\begin{equation}
    \frac{\mathrm{d}}{\mathrm{d}\lambda} \mathcal{L}_{\lambda} = \mathrm{det} \ T_{\mu\nu}^{(\lambda)} \ .
\end{equation}
The purpose of this section is to show what field-theoretic T-duality looks like for specific scaling limits of theories derived as $\ttb$ deformations of the $AB$-system. 

While chiral theories do not flow under a $\ttb$ deformation, a
theory with both left- and right-chiral bosons does undergo a
nontrivial deformation \cite{Chakrabarti:2020dhv}. The
$\ttb$-deformed Lagrangian was found to be
\begin{equation}
  \label{eq:ttb-deformed-chiral-bosons}
L = L _{A} + L _{B} + L _{V} \ ,
\end{equation}
where $ L _{B} $ is Sen's Lagrangian for an anti-chiral $ 1 $-form
$ B $. We also have the contribution arising from the $\ttb$
deformation
\begin{equation}
  \label{eq:ttb-deformed-piece}
L _{V} = \frac{1}{2 \lambda } \left[ 1-\sqrt{1+V^2} + V \sinh ^{-1} V  \right] \ ,
\end{equation}
where $ \lambda $ is the $\ttb$ coupling constant and $ V $ is the
dimensionless combination\footnote{In \cite{Chakrabarti:2020dhv} the
  Lagrangian was expressed in terms of the variable $X = V^2$. Here,
  we use $V$ since it will make our expressions simpler.}
\begin{equation}
  \label{eq:what-is-v}
V = \lambda  \left(A \wedge \star B\right) \ .
\end{equation}
In the limit $ \lambda \rightarrow 0 $, one recovers the original
theory.

We now consider the following limit:
\begin{equation}
  \label{eq:decoupling-limit}
\lambda \rightarrow \infty \quad \mathrm{with} \quad V \quad \mathrm{fixed.}
\end{equation}
In this limit, the piece $ L_{V} $ from the $\ttb$ deformation in
\cref{eq:ttb-deformed-chiral-bosons} is negligible and one recovers
decoupled left- and right-chiral bosons. This is consonant with
expectations that infinitely strongly coupled $\ttb$-deformed
theories are chirally decoupled \cite{Chakrabarti:2020pxr}.

Emboldened by this observation, we now consider the opposite limit,
where
\begin{equation}
  \label{eq:new-limit}
\lambda \rightarrow 0 \quad \mathrm{with} \quad V \quad \mathrm{fixed.}
\end{equation}
The term $ L_{V} $ is dominant in this limit. The reader may worry
that such a limit is ill-defined on account of the auxiliary scalar
$ \phi $ in $ L _{A} $. Consider the field rescalings
\begin{equation}
\phi \rightarrow \sqrt{\lambda } \phi \quad \mathrm{and} \quad A \rightarrow \frac{1}{\sqrt{\lambda } } A \ ,
\end{equation}
that effect the limit prescribed in \cref{eq:new-limit}. This sends
\begin{equation}
L _{A} \rightarrow \frac{\lambda }{2} \mathrm{d} \phi \wedge \star \mathrm{d} \phi - 2 A \wedge \mathrm{d} \phi \ ,
\end{equation}
and so in the limit $ \lambda \rightarrow 0 $ the kinetic term for the
auxiliary scalar drops out, and we are left with just the second
term. More precisely, we are left with the wedge product of $ A $ with
the self-dual part of $ \mathrm{d} \phi $, courtesy of the
self-duality of $ A $. On using the fact that we can discard boundary
terms in the corresponding action, meaning we can replace
\begin{equation}
\int _{} ^{} \mathrm{d} A \wedge \phi = \int _{} ^{}  A \wedge \mathrm{d} \phi \ ,
\end{equation}
we are left with
\begin{equation}
-A \wedge \left( \mathrm{d} \phi + \star \mathrm{d} \phi  \right) = -\phi \wedge \mathrm{d} \left( A - \star A \right) \ ,
\end{equation}
which is identically zero off-shell due to the self-duality of $ A
$. (A similar exercise for $ L _{B} $ results in the same conclusion,
this time courtesy of the anti-self-duality of $ B $.)  To summarise,
the limit in \cref{eq:new-limit} is well-defined even in the quantum
theory since the path integral over the auxiliary scalar field in this
limit is a Lagrange multiplier enforcing a trivial constraint. Said
differently, $ L_{V} $ really is dominant in this regime and
$ L _{A} $ and $ L _{B} $ can be safely neglected.

Furthermore, this limit can be interpreted (perhaps paradoxically!) as
an ultraviolet limit where the effect of the undeformed theory fades
away, and the deformation dominates. In this limit, we expect the
ultraviolet degrees of freedom to be carried out solely by the field
$V$ itself and not the constituent chiral form fields $A$ and $B$,
which were fundamental in the infrared regime. Here onwards, we
consider the field $V$ to be fundamental whenever we discuss physics
in this novel scaling limit.

The Lagrangian $L_V$, after a dimensional reduction on a circle of
radius $r'$, must lead to a quantum mechanical system. It
is straightforwardly seen that the effective one-dimensional Lagrangian
(ignoring a ``cosmological constant'' term) is given by
\begin{equation}
  \label{eq:L-V-1d}
    L_{V} = \frac{2 \pi r'}{2 \lambda}\left[ - \sqrt{1+v_i v_{-i}} + \sum_{N} c_N \, v_{i_1} v_{i_2} \cdots v_{-(i_1+i_2+ \cdots +i_{2N-1})} \right] \, ,
\end{equation}
where $v_i$ are the modes in the expansion of $V(x,t)$ along the circle directions
\begin{equation}
V(x,t) = \sum_{n}^{} v _{n} \, \mathrm{exp}\left\lbrace \frac{\mathrm{i}n x}{r}  \right\rbrace \ .
\end{equation}
Note that we have expressed $V \sinh^{-1} V = \sum_N c_N V^{2N}$ for
appropriate coefficients $ c _{N} $. Strictly speaking, an additional
contribution is also present because the metric with a circle
direction is no longer flat, and Sen's action on a curved background
receives additional contributions. In what follows, however, we will
eventually always return to the flat-space (or decompactification)
limit; therefore, this additional term will not play any role in our
analysis.

Also note, as pointed out in \cref{sec:sugra}, in the absence of the kinetic pieces, we can no longer suppress the contributions from the higher modes, even at the limit $r \to 0$. Indeed, this further justifies our choice to treat the field $V(x,t)$ as fundamental in this novel scaling limit.

One of the key observations of this section is that the deformed theory
in this novel scaling limit is T-dual to a particular deformation of
quantum mechanics with a non-standard kinetic term that has already
appeared in the literature, which we will refer to as the
Grassi-Mari\~no (GM) deformation of quantum mechanics
\cite{Grassi:2018bci}. The remainder of this section will be dedicated to
explicitly establishing the duality map and discussing the
consequences of such an identification.

\subsection{Connection to T-Duality}
The deformation considered in \cite{Grassi:2018bci}, in the absence
of a potential, corresponds to a Hamiltonian
\begin{equation}
    H_{\mathrm{GM}} = 2 \cosh{p} \ .
\end{equation}
The Hamiltonian, as defined above, is dimensionless (as is the
momentum variable $p$). One can easily find the Lagrangian
corresponding to this deformed quantum mechanics, which will also be
dimensionless. Typically, the Lagrangian has unit mass dimension in
one dimension and natural units. To ensure this, we define
\begin{equation}
  \label{eq:gm-lagrangian}
    L_{\mathrm{GM}} =  \frac{1}{2 \pi r} \left [ - \sqrt{1+v^2} + v \sinh^{-1} v \right] \ ,
\end{equation}
for some length scale $ R $. We have deliberately denoted the velocity
variable $\dot{q}$ by $v$, anticipating the identification with the
effective one-dimensional theory obtained in \cref{eq:L-V-1d}. Our
task now is to relate two theories: the first
$ \left( L _{\mathrm{GM}} \right) $ is a deformation of ordinary
quantum mechanics, and the second $ \left( L_{V} \right) $ arises in
the context of integrable deformations of a specific two-dimensional
quantum field theory. We will do so using a glueing prescription
motivated by the study of field-theoretic T-dualities
\cite{Taylor:1996ik,Ishii:2007sy,Yamazaki:2019prm}.

Consider an infinite number of copies of this theory, each defined on
a circle of radius $R$.\footnote{Since
  the only dimensionful parameter of $L_{\mathrm{GM}}$ was a length
  scale $R$, it makes sense to interpret it as the radius a circle
  direction along which this theory should be T-dualised.} We promote
the variable $v$ to an infinite number of variables $v_{i,j}$ which
denote the winding modes of $v$ from $ i ^{\mathrm{th}} $ copy to the
$ j ^{\mathrm{th}} $ copy with winding number $(i-j)$. We impose an
overall periodicity that ensures the Lagrangian in the covering space
involves a trace over the circle indices.\footnote{The sum over
  repeated circle indices is implicit.}  This yields
\begin{equation}
L^{\infty}_{\mathrm{GM}} = \frac{1}{2 \pi r} \left[ - \sqrt{1+v_{i,j}v_{j,i}} + \sum_{N} c_N \, v_{i_1,i_2} v_{i_2,i_3} \cdots v_{i_{2N},i_1} \right] \ .
\end{equation}

Next, we identify the field variables along successive circles:
$v_{i+1,j+1} = v_{i,j}$. This essentially allows us to use $v_{m,0}$
as the sole independent degree of freedom. In what follows, we use the
notation $v_m = v_{m,0}$. The Lagrangian
$ L _{\mathrm{GM}}^{\infty } $ simplifies to
\begin{equation}
  \label{eq:L-infinity-gm}
L^{\infty}_{\mathrm{GM}} = \frac{1}{2 \pi r} \left[ - \sqrt{1+v_i v_{-i}} + \sum_{N} c_N \, v_{i_1} v_{i_2} \cdots v_{-(i_1+i_2 +\cdots +i_{2N-1})} \right] \ .
\end{equation}

At this stage, we are ready to compare $L^{\infty}_{\mathrm{GM}}$ in
\cref{eq:L-infinity-gm} with $ L_V$ in
\cref{eq:L-V-1d}.  Noting that $\lambda \sim \ell _s^2$ for some
length scale $\ell _s$, we observe that we have the same theory
provided
\begin{equation}
    \frac{1}{2 \pi r} = \frac{2 \pi r'}{2 \lambda} = \frac{\pi r'}{\ell _s^2} \ .
\end{equation}
The correspondence between the two is plainly a T-duality relation.

\begin{figure}
    \centering
    \includegraphics[width=1\linewidth]{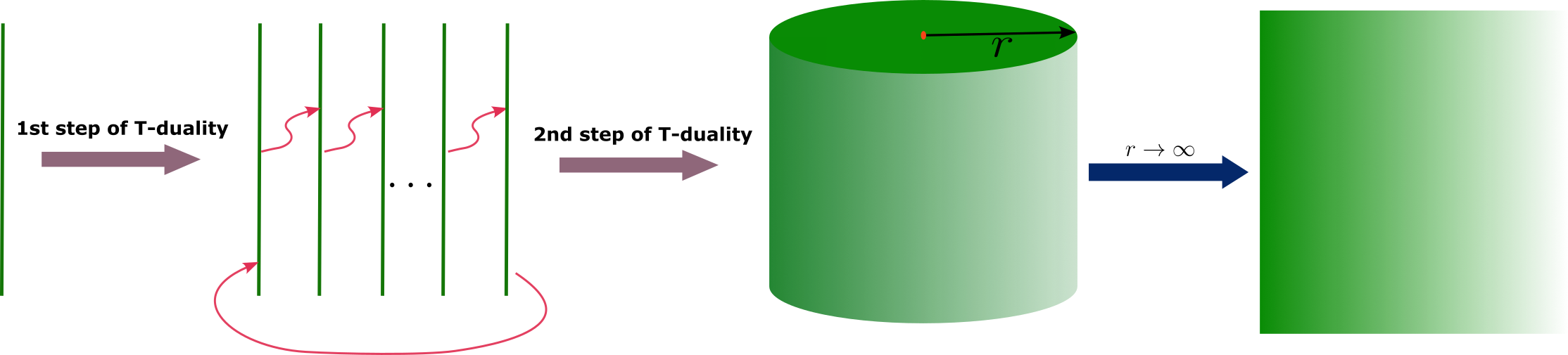}
    \caption{From Grassi-Mari\~no QM to $2D$ $\ttb$-deformed chiral bosons in the novel scaling limit. The entire process involves first a field-theoretic T-duality which uplifts the QM on a line to a QFT on a cylinder. Finally, decompactifying the cylinder to a plane recovers the desired answer. }
    \label{fig:full-T-duality}
\end{figure}

Eventually, of course, we would like to take the decompactification
limit $r' \to \infty$ in \cref{eq:L-V-1d}. Notice that this
decompactification limit is essentially identical to the limit
$\lambda \to 0$ with $V$ held fixed, which we have already established
isolates $ L_{V} $ from the entire $\ttb$-deformed
Lagrangian. Conversely, the limit $ r' \rightarrow 0 $ is equivalent
to the limit $ \lambda \rightarrow \infty $ at fixed $ V $, which we
have already seen recovers the familiar free dynamics of chiral
fields. The whole process is summarized in the \Cref{fig:full-T-duality}.

We therefore conclude that the Grass-Mari\~no Lagrangian, T-dualised
in the aforementioned manner, when decompactified is equivalent to the
scaling limit in \cref{eq:new-limit} of the $\ttb$-deformed left- and
right-chiral bosons in Sen's formalism. This identification offers a
different perspective on both the nature of and the integrability
properties enjoyed by the Grassi-Mari\~no deformation: it descends
from the integrability of the $\ttb$ deformation in one higher
dimension.

Of course, given the free Hamiltonian $H = p^2$, the deformed
Hamiltonian $H = 2 \cosh{p}$ is an example of a quantum mechanical
analogue of the $\ttb$ deformation in the sense of
\cite{Gross:2019uxi}. Therefore, even before this T-duality map, one
could anticipate a relation between the Grassi-Mari\~no and $\ttb$
deformations. However, in this specific instance, we are able to do
one better, as it were: our analysis allows us to explicitly identify
a parent $\ttb$-deformed two-dimensional theory.

\textbf{Dimensional Reduction.}
Given the similarity between the Lagrangians
\cref{eq:ttb-deformed-piece} and \cref{eq:gm-lagrangian}, the reader
may wonder: isn't the relation between the two just dimensional
reduction on a circle? Alternatively, is there a parametrisation of the ultraviolet degrees of freedom $(V)$ such that on dimensional reduction, it goes into the Grassi-Mari\~no Lagrangian?

It is easy to verify that the only choice of parametrisation that does the job is
\begin{equation}
\label{eq:v2-parametrisation}
V^{2} = \lambda \, \mathrm{d} \varphi \wedge \star \mathrm{d} \varphi \ ,
\end{equation}
With this parametrisation, after dimensionally reducing on a circle of vanishingly
small radius $ r' $, we observe that the modes $ \varphi _{n>0} $ on
the circle are heavy and therefore at low energies only the dynamics
of the zero mode $\varphi_{0}$ survives, sending $ L _{V} $ to 
\begin{equation}
L _{\varphi } = \frac{2 \pi r'}{2 \lambda } \left[ -\sqrt{1+\dot{\varphi }_{0} ^{2}} + \dot{\varphi }_{0} \sinh ^{-1} \dot{\varphi }_{0}  \right] \ ,
\end{equation}
which is the same as \cref{eq:gm-lagrangian} on identifying
$ q $ with $ \varphi _{0} $ and $ r \sim \lambda / r' $.

While the two Lagrangians $L_V$ and $L_{\varphi}$ are same given the identification in \cref{eq:v2-parametrisation}, the two give rise to very different dynamics when thought of as variational problems. That is, the identification in \cref{eq:v2-parametrisation} is acceptable in principle, but no longer admits the interpretation of a $\ttb$-deformed theory of free chiral and anti-chiral bosons. Throughout this discussion, we have worked within Sen's formalism and our results crucially depend on our working with fundamental field strength-like variables and not invoking the corresponding gauge potentials.

Of course, we would be remiss if we did not point out that the
discussion in this section is restricted to the kinetic part of the
Grassi-Mari\~no spectral problem. We now turn on the potential and see what are its effects on the two-dimensional QFT.

\subsection{Including Potentials}
If one considers a Hamiltonian $H = 2 \cosh{p} + W(q)$, one can then
go on to find its T-dual following the prescription outlined above. We
no longer have freedom to choose how to glue the coordinates, since
$v = \dot{q}$ and the glueing rules imposed on $v_m$ naturally impose a
glueing rule for the coordinates $q_m$.

Assuming a polynomial potential $W(q) = \sum_N g_N q^N$ for real
dimensionful coefficients $g_N$ such that $V(q)$ has unit mass
dimension in natural units. This leads to
\begin{equation}
    W(q_m) = \sum_N g_N q_{i_1} \cdots q_{-(i_1 + \cdots + i_{N-1})} \ ,
\end{equation}
where the sum over the circle indices is once again implied.

These can be identified with the mode expansion of the following two-dimensional
function on the circle direction
\begin{equation}
\begin{aligned}
    \int^t \mathrm{d} \tau \, V(x , \tau) &= \int^t \mathrm{d} \tau  \sum _{N} v_N(\tau) \exp{\left\lbrace\frac{i N x}{r}\right\rbrace} \ ,  \\
                             &= \int^t \mathrm{d} \tau \sum _{N} \dot{q}_N(\tau) \exp{\left\lbrace\frac{i N x}{r}\right\rbrace} \ , \\
                             &= \sum _{N} q_N(\tau) \exp{\left\lbrace\frac{i N x}{r}\right\rbrace} \ .
\end{aligned}
\end{equation}
Therefore, the two-dimensional term in the Lagrangian that is T-dual
to a quantum mechanical potential looks like
\begin{equation}
    W(V) = \sum_N g_N \prod_{i=1}^N \int^t \mathrm{d} \tau_i \, V(x, \tau_i) \ .
\end{equation}
This potential term is immediately seen to contain two unusual
features, which we now discuss.

\textbf{Locality.}
The potential term appears to be non-local in time. While
this may seem unpalatable at first sight, it is not unexpected. Even the familiar Floreanini-Jackiw formalism for chiral bosons is actually a non-local action when expressed in terms of local fields \cite{Floreanini:1987as,Sonnenschein:1988ug}. Furthermore, even without the potential term, the Lagrangian $L_V$ is non-local, as expected from any $\ttb$-deformed bosonic theories. Finally, recall that $\int A$ and $\int B$ appeared naturally in our attempt to define a vertex operator. Therefore, it stands to reason that while the terms in the potential naively appear to be non-local, they might not be so. Without a specific choice of a potential function, any further comments on purely general ground is not feasible.

\textbf{Lorentz Invariance} The potential term will typically break
Lorentz invariance. This is apparent since the integration that
appears in the potential term only involves the temporal
coordinate. Once again, this is to be expected. A Hamiltonian of the
form $H = 2 \cosh p + V(q)$ cannot be generically expected to descend
from a Lorentz invariant quantum field theory! In fact, it is quite
remarkable that the momentum-dependent piece \emph{does} descend from
a relativistic field theory. A non-relativistic quantum mechanics may
descend from a relativistic field theory as long as some subtle
aspects are treated with due diligence. These are beyond the scope of
this article, but the interested reader may wish to consult
\cite{Padmanabhan:2017bll} for more detailed discussions.

\section{Future Directions} \label{sec:discussion}

The main goal of this paper was to explore different aspects of T-duality symmetry of theories involving self-dual field strengths. For various reasons, but mostly because of its origin in string field theory, we chose Sen's formalism to capture the dynamics of the self-dual field strength. We have discussed three avatars of T-duality in this paper: one relating Type II supergravities, one on the worldsheet, and one in field theory. Since each of these sections included fairly extensive commentary, we restrict ourselves here to a discussion of the future directions naturally suggested by the results we have presented.

On the supergravity and worldsheet fronts, the principal direction in which our analyses can be extended is to consider more general dynamical target spacetimes. This is not just of interest from a technical standpoint; our contention is that a synthesis of Sen's formalism with doubled worldsheet and doubled field theory approaches should lead to new perspectives on the bottom-up construction of manifestly T-duality invariant spacetime actions, which are low-energy effective descriptions of string theory. In addition, it would be fascinating to see how Sen's formalism can capture non-geometric backgrounds and non-Abelian T-duality. Similar arguments were also presented, at least in spirit, albeit from a different perspective, in \citep{Hull:2023dgp}.

A major observation that emerged from this paper is that it is worthwhile to investigate how to rewrite all RR $p$-form field strengths in Sen's formalism. Given the general predictions based on superstring perturbation theory \citep{Witten:2012bh}, this rewriting to achieve a fresh formulation of supergravity is truly needed. It will be especially interesting to see how the additional field joins the fray and what subtle roles they play \citep{Lambert:2023qgs}. Furthermore, it has long been recognized that the quantisation of self-dual fluxes and how they change under dualities contain various nuances \citep{Witten:1999vg}, and that RR fields are globally classified by K-theory \citep{Minasian:1997mm,Moore:1999gb,Freed:2000tt}. It would be fascinating to examine how Sen's formalism relates to these broad global results for self-dual field strengths.

We have also presented new results regarding a field-theoretic T-duality of $\ttb$-deformed chiral bosons in Sen's formalism. It is curious that the scaling limit we defined does not appear to lead to a sensible theory for any other $\ttb$-deformed bosonic actions; it seems that this conclusion is unique to the theory considered here and crucially depends on the fact that Sen's formalism was used to construct the deformed Lagrangian. 

We saw that the resulting theory in this novel scaling limit is related by T-duality to a known integrable deformation of quantum mechanics first described in \cite{Grassi:2018bci}, and concluded that perhaps only the kinetic piece of the deformed quantum mechanics makes
rigorous sense in this T-dual map. Nevertheless, this
highlights an essential point regarding Sen's formalism. In cases
where the field theory of self-dual fields has a stringy origin,
adopting an approach where field strength variables are fundamental
makes perfect sense --- this is uncontroversial. However, as shown in
\cite{Sen:2019qit}, the formalism is equally well suited to handling
chiral bosons (and other self-dual form fields) without any stringy
embedding. In all such cases, the interactions one can introduce in
Sen's formalism seem to be in stark contrast with other approaches to
chiral bosons where gauge potentials continue to be used.\footnote{See
  \cite{Evnin:2022kqn} for a recent comparison between various leading
approaches.} In order to make contact with other approaches, all of
which should describe the same physics, one needs to introduce
flux-like variables in Sen's formalism that \emph{prima facie} appear
non-local. We hope to return to a systematic investigation of this
question in a future project.

The analysis of this article sheds light on certain aspects of the
$\ttb$-deformed theory of \cite{Chakrabarti:2020dhv} that were not
entirely clear before. It was suggested then that just as the $\ttb$
flow connects the free (ordinary) boson to the Nambu-Goto string, our
non-local action too might admit a target space interpretation. A
T-duality map provides compelling evidence in favor of this
suggestion. It would be exciting to understand the dynamics of the
extended object that leads to this deformed theory.

Further, the deformed Lagrangian obtained in
\cite{Chakrabarti:2020dhv} is classical. It was used to perform the
first explicit quantum scattering computation within Sen's formalism
in \cite{Chakrabarti:2022lnn}, where we verified that the deformed
quantum theory continues to be integrable up to $1$-loop. A full proof
of quantum integrability remained out of reach given the
limitations inherent in perturbation theory. The T-duality map derived
here suggests that the two-dimensional theory is quantum integrable in
a different region of the parameter space that was not accessible to
the standard perturbation theory. While this demonstration is by no
means a proof, it significantly strengthens the expectation that
$\ttb$-deformed chiral bosons in Sen's formalism also remains quantum
integrable.

The connection of $\ttb$-deformation and T-duality has appeared
before, albeit in a very different context. In \cite{Araujo:2018rho},
it was argued that $\ttb$-deformed conformal field theories are
holographic dual to TsT transformations of NS-NS backgrounds of
strings on $\mathrm{AdS}_{3} \times S^3 \times T^4$. (See also \cite{Apolo:2019zai} for the case of $\mathrm{BTZ} \times S^3 \times M^4$.) It is not \emph{a
  priori} clear if these two are at all related. It does, however,
open up the possibility that through a series of duality maps
(involving both T-duality and holography), the Grassi-Mari\~no
deformations of quantum mechanics can possibly capture particular
string backgrounds, which makes an already curious deformation of
quantum mechanics even more intriguing.

\subsection*{Acknowledgements}
The authors are grateful to Sujay Ashok, Adwait Gaikwad, Renann Lipinski Jusinskas, and Horatiu Nastase for discussions. MR is supported by an Inspire Faculty Fellowship.

	\bibliography{refs}
\end{document}